\documentclass{article}
\usepackage{amsmath}
\usepackage[latin1]{inputenc}

\newcommand{\Ket}[1]{\left\vert#1\right\rangle}
\newcommand{\Bra}[1]{\left\langle#1\right\vert}

\begin{document}

\title{Geometric phase accumulation-based effects in the
quantum dynamics of an anisotropically trapped ion}

\author{
M. Scala\footnote{email:matteo.scala@fisica.unipa.it}, B.
Militello and A. Messina\\
\small\it INFM, MIUR and Dipartimento di Scienze Fisiche ed
Astronomiche\\
\small\it dell'Università degli Studi di Palermo,
via Archirafi 36, I-90123 Palermo, Italy\\
}

\date{ }

\maketitle

\begin{abstract}
New physical effects in the dynamics of an ion confined in an
anisotropic two-dimensional Paul trap are reported. The link
between the occurrence of such manifestations and the accumulation
of geometric phase stemming from the intrinsic or controlled lack
of symmetry in the trap is brought to light. The possibility of
observing in laboratory these anisotropy-based phenomena is
briefly discussed.
\end{abstract}

keywords: Berry phase, Ion Traps\\

PACS: 39.10.+j, 03.65.Vf

\section{Introduction}
The continuous development of sophisticated techniques of cooling
and trapping atoms witnessed over the last years provides physical
systems ideal to test fundamental aspects of quantum mechanics. In
particular ions confined in a Paul trap, an appropriately designed
configuration of oscillating electromagnetic fields, exhibit a
centre of mass (c.m.) motion describable as that of a quantum
harmonic oscillator \cite{ghosh,nist,nist2}. With the help of a
classical laser field it is possible to induce in the ion
couplings between its vibrational and electronic degrees of
freedom representable by more or less simple spin-boson
hamiltonian models. When the laser wavelength is much larger than
the amplitude of the ion c.m. oscillations (Lamb-Dicke regime) and
the frequency of the classical field is tuned to one of the
vibrational sidebands of the electronic transition under scrutiny
(two-level approximation), the effective interaction may be
described by Jaynes-Cummings-like models
\cite{vogel,vogel1,messina,sabrinaPRA,danielePRA}.

The use of ion traps instead of cavities to test the dynamics
originating from hamiltonian models of this kind has the advantage
that environment effects on the ion may be reasonably neglected
\cite{nist}.

In this paper our attention is focused on {\em anisotropic}
two-dimensional Paul traps, that is characterized by different
oscillation frequencies along the principal axes of the trap. Our
scope is to bring to light effects which are directly traceable
back to such a lack of symmetry in the trap. When the ion is
driven by a laser beam not collinear with any of the axes of the
trap, we show that, unlike what happens in the isotropic case,
there is no way to describe the system dynamics in terms of a
time-independent interaction picture hamiltonian. We instead prove
the existence of a simple and physically transparent resonance
condition under which the interaction picture hamiltonian becomes
sinusoidally oscillating at the difference between the two free
oscillation frequencies of the c.m. of the ion. This implies that
when the anisotropy is small enough the system dynamics may be
faced with using the adiabatic approximation approach. Therefore
the system we are going to study proves to be an ideal candidate
to seek effects stemming from the Berry or geometric phase
accumulation \cite{berry,berry2}.

The main result reported in this paper is twofold. On the one hand
we show that anisotropy can be considered as a valuable resource
to test the theory of adiabatic evolutions and non-dynamical
phases.

On the other hand we put into evidence that the feasibility of
experiments for measuring the Berry phase provides a way to reveal
and quantitatively appreciate the presence of anisotropy in the
trap.

The paper is organized as follows. In the next section we describe
the physical system under study and in the third section its
time-dependent effective hamiltonian under the action of an
appropriate laser field is obtained. In the fourth section we
analyze the adiabatic evolution of the physical system and
calculate the Berry phase acquired by the instantaneous
eigenstates of the hamiltonian. The fifth section is devoted to
physical effects directly attributable to geometric phase
accumulation. Finally in the sixth section some conclusive remarks
are given.

\section{Ions in a Paul trap}
A Paul trap is a configuration of electrodes which generates a
quadrupolar electromagnetic potential $\Phi(t)$ that, using
cylindrical coordinates $r$, $\phi$, $z$, can be written as
\cite{ghosh,nist,nist2}
\begin{equation}\label{cap1phistatico}
  \Phi(t)=\Phi_0(t)\frac{(r^2-2z^2)}{2r_0^2}
\end{equation}
where $r_0$ is a length parameter (trap radius) depending on the
dimensions of the trap and $\Phi_0(t)$ introduces an appropriate
time dependence.

If $\Phi_0(t)=U$ is constant, i.e. if the field generated by the
electrodes is electrostatic, we cannot obtain any confinement for
a charged particle moving in such a field, as well known from
classical electromagnetic field theory.

Confinement of a charged particle may be easily achieved by
superposing to the static potential $U$ an oscillating one having
amplitude $V$ and frequency $\Omega$:
\begin{equation}\label{cap1potosc}
 \Phi_0(t)=U-V\cos(\Omega t)
\end{equation}

In this case, putting
\begin{equation}\label{cap1pagparam}
a_z=-2a_r=-\frac{8eU}{mr_0^2\Omega^2}\,,\;\;\;\;
  q_z=-2q_r=-\frac{4eV}{mr_0^2\Omega^2}\,,\;\;\;\;
  \zeta=\frac{\Omega t}{2}
\end{equation}
the equations of motion for a particle with charge $q$ and mass
$m$ admit stable solutions, i.e. confinement for the charged
particle, if the values of the parameters $a_z$, $q_z$, belong to
appropriate zones of the space of parameters called {\em stability
zones} \cite{ghosh}. Moreover, in this case it is possible to
separate the charged particle motion into the sum of two
components, one of which, called {\em micromotion}, is a very
rapid oscillation with an amplitude small enough to be negligible
with respect to the main component, the {\em secular motion}. The
latter is a harmonic motion corresponding to an effective
three-dimensional harmonic potential whose frequencies are given
by \cite{ghosh}:
\begin{equation}\label{frequenze}
  \omega_r^2=\left(a_r+\frac{q_r^2}{2}\right)\frac{\Omega^2}{4}\,,\;\;\;\;
  \omega_z^2=\left(a_z+\frac{q_z^2}{2}\right)\frac{\Omega^2}{4}\\
\end{equation}
and with which we can replace the real time-dependent potential
given by eqs.(\ref{cap1phistatico}) and (\ref{cap1potosc}).

It is important to emphasize that if we choose $U$ and $V$ in such
a way that
\begin{equation}\label{con_pot_sferico}
 U=\frac{eV^2}{mr_0^2\Omega^2}
\end{equation}
we obtain $\omega_r=\omega_z$ and so the quadrupolar potential
corresponding to these values gives rise to an effective potential
which is spherically symmetric. Thus, starting from values of $U$
and $V$ related in accordance with eq. (\ref{con_pot_sferico}),
changing at will the static potential $U$ allows the introduction
of anisotropy in a prefixed axial plane (for instance $x-z$) of
the Paul trap.

In particular we claim that current technology is compatible with
the possibility of designing and realizing small Paul traps
wherein a weak fully controllable anisotropy may be established.

Of course  even an imperfect construction of the electrodes, for
instance a bad or moderately good realization of their required
geometric properties, may lead to an experimental setup operating
in non-ideal conditions modelizable in terms of anisotropy, for
example in the radial plane $(x-y)$.

\section{Physical system and hamiltonian model}
Let us consider a two level ion of mass $m$ confined in a
two-dimensional $x-z$ Paul trap, whose oscillation frequencies are
$\nu_a$ and $\nu_b$ respectively, interacting with a laser beam
oriented along a direction which makes an angle $\alpha$ with the
$x$ axis. It has been shown that this system, in the rotating wave
approximation, may be described by the following hamiltonian model
\cite{vogel}:
\begin{equation}\label{hamilt_schr}
  H_{RW\!\!A}=\hbar\left(\nu_a a^\dag a+\nu_b b^\dag
  b\right)+\frac{\hbar\omega_0}{2}\sigma_z+\left[d\epsilon^{(-)}\left(x,z,t\right)\sigma_-+h.c.\right]
\end{equation}
where $d$ is the appropriate dipole transition matrix element,
$\hbar\omega_0$ the energy separation between the ground
($\Ket{-}$) and the excited ($\Ket{+}$) electronic states of the
ion and $\sigma_z=\Ket{+}\Bra{+}-\Ket{-}\Bra{-}$,
$\sigma_+=\Ket{+}\Bra{-}$ and $\sigma_-=\Ket{-}\Bra{+}$.

The quantity $\epsilon^{(-)}\left(x,z,t\right)$ is the negative
frequency part of the classical driving field, given by:
\begin{equation}\label{negative_part}
 \epsilon^{(-)}\left(x,z,t\right)=E_0\,\mathrm{e}^{-i\left[k\left(x\cos\alpha+z\sin\alpha\right)-\omega t\right]}
\end{equation}
where $\omega$ and $k$ are the frequency and the modulus of the
wave vector of the laser respectively.

In a frame rotating at the laser frequency, the hamiltonian can be
cast in the form $H=H_{trap}+H_{int}$, with:
\begin{equation}\label{cap4schema_ad_uno}
 \left\{\begin{array}{l}H_{trap}=\hbar\left(\nu_a a^\dag a+\nu_b b^\dag b\right)+\hbar\frac{\delta}{2}\sigma_z\\
   \\
   H_{int}=\hbar\Omega\left\{e^{-ik\left(x\cos\alpha+z\sin\alpha\right)}\sigma_-+
   e^{ik\left(x\cos\alpha+z\sin\alpha\right)}\sigma_+\right\}
 \end{array}\right.
\end{equation}
where $\delta=\omega_0-\omega$ is the detuning of the laser beam
from the electronic transition frequency,
$\Omega=\frac{dE_0}{\hbar}$ and $a$ ($a^\dag$) and $b$ ($b^\dag$)
are the annihilation (creation) operators along the directions $x$
and $z$ respectively.

In the interaction picture with respect to the hamiltonian
$H_{trap}$ given by (\ref{cap4schema_ad_uno}), the exponentials
appearing in $H_{int}$ can be expanded to obtain a sum of terms
which oscillate at frequencies which are linear combinations of
$\nu_{a}$, $\nu_{b}$ and ${\delta}$ with integer coefficients.

In the Lamb-Dicke limit $k\Delta x\cos \alpha , \;k\Delta
z\sin\alpha <<1$ (where $\Delta x=\sqrt {\frac{\hbar}{2m\nu
_a}}\,$ and $\Delta z=\sqrt{\frac{\hbar}{2m\nu_b}}$), under the
simple {\em resonance condition} $\delta=\nu_a+\nu_b$ and assuming
$\Delta\nu\equiv\nu_a-\nu_b\ll\mathrm{min}\left\{\nu_a,\nu_b\right\}$,
the only terms which are both non negligible and slowly
oscillating are those proportional to ${a^\dag}^2\sigma_{-}$,
${b^\dag}^2\sigma_{-}$, $a^{\dag} b^{\dag }\sigma_{-}$ and their
hermitian conjugates. Performing the rotating wave approximation,
also taking into account that
$\Omega\ll\mathrm{min}\left\{\nu_a,\nu_b\right\}$, yields the
following interaction picture hamiltonian
\begin{equation}\label{cap4H_int_sch}
   H_{int}^I(t)\approx\,-\hbar\lambda\left[\left(\cos\theta\,a^\dag\mathrm{e}^{-i\frac{\Delta\nu}{2}\,t}+
  \sin\theta\,b^\dag\mathrm{e}^{i\frac{\Delta\nu}{2}\,t}\right)^2\sigma_-+h.c.\right]
\end{equation}
where
\begin{equation}\label{parametri}
  \left\{\begin{array}{l}\cos\theta=\frac{\Delta x\cos\alpha}{\sqrt{\left(\Delta x\cos\alpha\right)^2+\left(\Delta z
   \sin\alpha\right)^2}}\;,\,\;\;
   \sin\theta=\frac{\Delta z\sin\alpha}{\sqrt{\left(\Delta
   x\cos\alpha\right)^2+\left(\Delta z
   \sin\alpha\right)^2}}\;,\\
   \\
   \lambda=\frac{\Omega}{2}\;\mathrm{e}^{k^2\left(\Delta x^2\cos^2\alpha
+\;\Delta z^2\sin^2\alpha\right)}\,\left[\left(\Delta
   x\cos\alpha\right)^2+\left(\Delta
   z\sin\alpha\right)^2\right]\;.
 \end{array}\right.\,\,
\end{equation}
The quantity $\Delta\nu$ measures the anisotropy degree
characterizing the vibrational ion dynamics.

The physically transparent result we obtain is the periodically
time-dependent hamiltonian $H_{int}^I(t)$, following from the
chosen resonance condition. This periodicity suggests that the
system under scrutiny could be advantageously exploited to
investigate effects related to Berry phase. It is worth noting
that, had we chosen a different value of $\delta$, we would anyway
have obtained a time-dependent hamiltonian, which means that {\em
such an explicit temporal dependence in $H_{int}^I(t)$, periodic
or not, entirely stems from the presence of anisotropy in the
trap}.

The anisotropy in the system under scrutiny gives thus rise at
resonance to an {\em intrinsic} periodicity of the hamiltonian and
to a possible accumulation of Berry phase in a period. With {\em
intrinsic} we mean that periodicity and Berry phase do not appear
as due to the manipulation of appropriate parameters, as, on the
contrary, we see in the well known case of spin 1/2 in an uniform
varying magnetic field \cite{berry}, but to the impossibility to
find a condition on the laser frequency such to wash out any
temporal dependence in the interaction picture hamiltonian.

In other words we conclude that the lack of symmetry of the trap
is at the origin of the appearance of Berry phase accumulation in
the system dynamics.

\section{Berry phase accumulation in a cycle}
\label{sec:calculation} It is well known that when the hamiltonian
$H(t)$ of a physical system is slowly varying with time, its
dynamical behaviour may be investigated with the help of the
so-called {\em adiabatic approximation} \cite{messiah}. Assuming
$H(t)$ possesses a non-degenerate discrete spectrum at any
arbitrarily fixed $t$, the time evolution of the $n$-th eigenstate
$\Ket{n(0)}$ of $H(0)$ generates, at the time instant $t$, the
$n$-th eigenstate $\Ket{n(t)}$ of $H(t)$ \cite{messiah}. The
anisotropy degree $\Delta\nu$ provides a useful reference
frequency to quantitatively estimate the rapidity of variation of
the instantaneous eigenstates of $H(t)$. Berry showed that there
is a simple way to express the phase acquired by $\Ket{n(t)}$
when, after some time $T$, the hamiltonian comes back to its
initial form, i.e. $H(T)=H(0)$.

Assume that the hamiltonian of the system depends on a set of
parameters ${\bf R}(t)=(R_{1}(t),R_{2}(t),...)$, i.e. $H(t)\equiv
H({\bf R}(t))$, and call $\{E_{n}({\bf R})\}$ the eigenvalues of
$H({\bf R})$ and $\{\Ket{n({\bf R})}\}$ the corresponding
eigenstates. The evolution of the hamiltonian $H(t)$ corresponds
to a curve in the space of parameters and the condition
${H(T)=H(0)}$ becomes ${\bf R}(T)={\bf R}(0)$.

If the initial state of the system is
$\left|\Psi(0)\right>=\left|n(0)\right>$, the state of the system
at the time instant $T$ may be written down as \cite{berry}:
\begin{equation}\label{cap2psiciclo}
 \left|\Psi(T)\right>=\mathrm{e}^{i\gamma_n^C}
 \mathrm{e}^{-\frac{i}{\hbar}\int_0^Tdt'\,E_n({\bf R}(t'))}
 \left|\Psi(0)\right>
\end{equation}
where the exponential factor containing the integral symbol,
called \textit{dynamical phase factor}, reduces to the usual phase
factor in the case of time-independent hamiltonian. The quantity
$\gamma _{n}^{C}$ is given by:
\begin{equation}\label{cap2gammaC}
 \gamma_n^C=i\oint_C\left<n({\bf R})\right.\left|\nabla\!_R
 \,n({\bf R})\right>\cdot {\bf dR}
\end{equation}
the curve $C$ being the path followed by ${\bf R}(t)$ in the space
of parameters. The quantity $\gamma_n^C$ is called {\em Berry
phase} or {\em geometric phase}, since its value depends only on
the curve $C$. Putting $\phi=\Delta\nu\,t/2$, the hamiltonian
(\ref{cap4H_int_sch}) may be cast in the following form:
\begin{equation}\label{cap4Hphi}
 \begin{split}
  H_{\phi}=-\hbar\lambda\left[{A_{\phi}^\dag}^2\sigma_-+A_{\phi}^2\sigma_+\right]
 \end{split}
\end{equation}
where
$A_{\phi}^\dag=\cos\theta\,\mathrm{e}^{-i\phi}\,a^\dag+\sin\theta\,
 \mathrm{e}^{i\phi}\,b^\dag$.

It is easy to verify that the creation operator
$B_{\phi}^\dag=\sin\theta\,\mathrm{e}^{i\phi}\,a^\dag+\cos\theta\,
 \mathrm{e}^{-i\phi}\,b^\dag$
corresponds to a vibrational mode orthogonal to that associated to
$A_{\phi }$.

Let us introduce the bimodal Fock states
\begin{equation}\label{cap4nphi}
 \left|N\right>_{\phi}\equiv\left|N\right>_{A_{\phi}}\left|0\right>_{B_{\phi}}=
 \frac{{A_{\phi}^\dag}^N}{\sqrt{N!}}\left|0\right>_a\left|0\right>_b
\end{equation}
$\left|0\right>_a$ ($\left|0\right>_b$) being the vacuum state of
the vibrational mode associated to $a$ ($b$). It is immediate to
check directly  using eq. (\ref{cap4Hphi}) that the states
$\left|0\right>_{\phi}\left|-\right>$ and
$\left|1\right>_{\phi}\left|-\right>$ are degenerate eigenstates
of $H_{\phi}$ with vanishing energy. The other eigenstates  of
$H_{\phi}$ are singlets of energy
\begin{equation}\label{cap4autovphi}
  E_{N,\pm}=\pm\hbar\lambda\sqrt{N(N-1)}\;.\\
\end{equation}
and can be cast in the following form:
\begin{equation}\label{cap4autophi}
  \left|\Psi_{N\pm}(\phi)\right>=\frac{1}{\sqrt{2}}\left(\left|N\right>_{\phi}\left|-\right>\mp
  \left|N-2\right>_{\phi}\left|+\right>\right)\\
\end{equation}
provided $N\ge 2$.

We see from eq. (\ref{cap4autovphi}) that the coupling constant
$\lambda$, at any time instant $t$, is of the order of the lowest
Bohr frequency associated to the instantaneous spectrum of $H(t)$.

Denoting by $U(t)$ the evolution operator relative to $H_{\phi}$,
the dynamics of the instantaneous eigenstates (\ref{cap4autophi})
after a cycle, in view of eq. (\ref{cap2psiciclo}), is given by
\begin{equation}\label{cap4evol_aut}
 U(T)\left|\Psi_{N\pm}(\phi=0)\right>=\mathrm{e}^{-i\frac{E_{N,\pm}}{\hbar}T}
 \mathrm{e}^{i\gamma_{_{N\pm}}(T)}\left|\Psi_{N\pm}(\phi=0)\right>\\
\end{equation}

The associated geometrical phase $\gamma_{_{N\pm}}(T)$ can be
calculated by performing the integration in eq.(\ref{cap2gammaC})
along the curve given by ${\bf R}=(\cos\phi,\sin\phi)$ and varying
$\phi$ from 0 to $2\pi$, obtaining the following explicit
expression:
\begin{equation}\label{cap4berry_ciclica_fine}
 \gamma_{_{N-}}(T)=\gamma_{_{N+}}(T)\equiv\gamma_{_N}(T)=-2\pi(2\sin^2\theta-1)N
\end{equation}
apart from terms independent from $N$, here omitted since they do
not give rise to observable effects.

The dependence of Berry phase on $N$ is the key building block for
the result we are going to present in the next section: it indeed
suggests that its occurrence may be revealed by preparing the
system in a superposition of Fock states with different values of
$N$.

Before closing this section we wish to write down a condition for
the validity of the adiabatic approximation in the case under
scrutiny. The evolution of the system can be considered as
adiabatic if its anisotropy degree is sufficiently small with
respect to the effective interaction strength $\lambda$. More
quantitatively, the square modulus of the ratio between the
frequency $\Delta\nu$, measuring the maximum rapidity of variation
of the set $\{\Ket{n(t)}\}$ of the instantaneous eigenstates of
$H(t)$, and the minimum Bohr frequency appearing in the
instantaneous spectrum of $H(t)$ has to be much smaller than unity
\cite{messiah}. In our case this condition becomes:
\begin{equation}\label{cap4_cond_ad}
 \left|\Delta\nu\right|^2\ll\left|\lambda\right|^2
\end{equation}

A quantitative estimation of the applicability of inequality
(\ref{cap4_cond_ad}) in our situation will be done in the last
section.
\section{Anisotropy vs. Berry phase effects}
\label{sec:anisotropy} In accordance with eq.
(\ref{cap4berry_ciclica_fine}) we now elaborate our strategy aimed
at finding some physical effects directly traceable back to Berry
phase accumulation and consequently to the presence of anisotropy
in a two-dimensional Paul trap.

Firstly we choose an initial state quite simple in structure,
experimentally feasible and appropriate to reveal Berry phase
accumulation. In view of eq. (\ref{cap4berry_ciclica_fine}), the
quantum superposition of two different eigenstates of $H(0)$
accumulating different Berry phases meets our request. After this
step we have to imagine a suitable physically transparent
observable sensitive to the non-dynamical phase acquired by the
eigenstates of $H(0)$ appearing in the initial state expansion.

To this end we propose to start with the following superposition
\begin{equation}\label{cap4stato_t0}
 \left|\Psi(t=0)\right>=\frac{1}{\sqrt{2}}\left(\left|N\right>_{\phi=0}+
 \left|N+1\right>_{\phi=0}\right)\left|-\right>\\
\end{equation}
which may be prepared by methods similar to that used to prepare
Fock states in the experiments described in \cite{meekhof}.

In view of eqs. (\ref{cap4autophi})-(\ref{cap4evol_aut}) and
(\ref{cap4berry_ciclica_fine}) the evolution of the state given by
eq. (\ref{cap4stato_t0}) under $H_{\phi}$ may be cast in the
following form:
\begin{equation}\label{cap4evoltpsiT}
 \begin{split}
  &\left|\Psi(t=T)\right>=\frac{1}{\sqrt{2}}\left(U(T)\left|N\right>_{\phi=0}\left|-\right>+
  U(T)\left|N+1\right>_{\phi=0}\left|-\right>\right)=\\
  \\
  &=\frac{1}{\sqrt{2}}\left\{\mathrm{e}^{i\gamma_{_N}(T)}
  \left[\cos\!\!\left(\!\!\frac{E_{N,+}}{\hbar}\,T\!\!\right)\!\left|N\right>_{\phi=0}\left|-\right>+
  i\sin\!\!\left(\!\!\frac{E_{N,+}}{\hbar}\,T\!\!\right)\!\left|N-2\right>_{\phi=0}\left|+\right>\right]+\right.\\
  \\
  &\left.+\mathrm{e}^{i\gamma_{_{N+1}}(T)}\!\!
  \left[\cos\!\!\left(\!\!\frac{E_{N+1,+}}{\hbar}\,T\!\!\right)\!\!\left|N+1\right>_{\phi=0}\!\left|-\right>\!+
  \!i\sin\!\!\left(\!\!\frac{E_{N+1,+}}{\hbar}\,T\!\!\right)\!\!\left|N-1\right>_{\phi=0}\!\left|+\right>\right]
  \right\}
 \end{split}
\end{equation}

As expected by construction, {\em this state turns out to be the
sum of two states having different geometrical phases}.

In accordance with our strategy, we must look for an observable
sensitive to Berry phase accumulation. A simple choice is
represented by the operator
\begin{equation}\label{cap4osserv}
 \hat{O}=\frac{A_{\phi=0}^\dag+A_{\phi=0}}{2}
\end{equation}
since it connects the two states of the superposition
(\ref{cap4evoltpsiT}). This operator has a clear physical meaning,
being proportional to the position operator of the vibrational
mode corresponding to $\phi=0$.

If we choose the direction $\alpha $ of the laser such that
$\theta=\pi $/6, we obtain the following mean value after a cycle
\begin{equation}\label{cap4O(T)_pisesti}
 \begin{split}
  &\left<\Psi(t=T)\right|\hat{O}_I\left|\Psi(t=T)\right>=
  \frac{1}{2}\Bigg\{\cos\left(\left[\gamma_{_N}(T)-\gamma_{_{N+1}}(T)\right]+\nu\,T\right)\times\\
      &\times\left[\cos\left(\frac{E_{N,+}}{\hbar}\,T\right)
  \,\cos\left(\frac{E_{N+1,+}}{\hbar}\,T\right)\sqrt{N+1}\,+\right.\\
    &+\left.\sin\left(\frac{E_{N,+}}{\hbar}\,T\right)
  \,\sin\left(\frac{E_{N+1,+}}{\hbar}\,T\right)\sqrt{N-1}\,\right]\Bigg\}\\
 \end{split}
\end{equation}
where $\hat{O}_I$ is the observable $\hat{O}$ in the interaction
picture and $\nu=\frac{\nu_a+\nu_b}{2}$\;.

In view of eq. (\ref{cap4berry_ciclica_fine}) and in
correspondence of $\theta=\pi /6$ we get
\begin{equation}\label{deltagamma}
 \gamma_{_N}(T)-\gamma_{_{N+1}}(T)=\pi
\end{equation}
It thus  turns out that the mean value of $\hat{O}$ given by eq.
(\ref{cap4O(T)_pisesti}) is the negative of what we would obtain
if the only phase factor acquired after a cycle by an eigenstate
of the hamiltonian $H_{\phi}$ were the dynamical one. Taking $N$
large enough, and for an appropriate value of $\lambda$, the
quantity between square brackets is much larger than unity and the
difference in sign we have found can be considered as {\em
macroscopic}.

Moreover, let us consider the instantaneous energy eigenvalues
given by eq. (\ref{cap4autovphi}). Since they are independent of
time, their structure is the same both in the case under scrutiny
and in the case of an ion trapped in an isotropic trap
($\nu_a=\nu_b$) and interacting with a laser beam oriented along
the direction $\theta$ and with a detuning $\delta$ equal to the
frequency of the trap. Thus we can say that in both cases
(isotropic and anisotropic) one finds the same dynamical phases
and that the only difference in the system dynamics is the
appearance of the geometric phase factor in the case of an
anisotropic trap.

Summing up, we claim the following result: {\em the macroscopic
difference in sign we have found is a direct manifestation of
anisotropy. It allows us to distinguish at a certain time instant
T between an isotropic trap and an anisotropic one having
oscillation frequencies such that} $\Delta \nu $=4$\pi $/$T$.

Such a behaviour also reflects the different structures of the
states of the ion at the time instant $T$. Indeed from eqs.
(\ref{cap4evoltpsiT}) and (\ref{deltagamma}) it is possible to
show that the states reached in the two cases (isotropic and
anisotropic) are superpositions of the same couple of orthogonal
states: {\em the difference between an anisotropic trap and an
isotropic one consists in the quantum phase difference of the two
states in the superposition}. It is of particular relevance that
the states of the ion at the time instant $T$ in the two
situations are orthogonal.

\section{Conclusive remarks}
\label{sec:conclusions} In this paper we have investigated the
quantum dynamics of an ion confined in a two-dimensional Paul trap
characterized by a difference $\Delta\nu$ between the oscillation
frequencies along its two principal axes. We have shown that when
a controllable or intrinsic anisotropy is weak, a laser beam
appropriately oriented gives rise to a vibronic coupling
representable, in the interaction picture, by a periodic
hamiltonian of frequency $\Delta\nu$.

If the coupling  strength compared with $\Delta\nu$ makes it
legitimate studying the dynamics of the system under the adiabatic
approximation, we find a $\left|n(0)\right>$-dependent and trap
anisotropy-based Berry phase accumulation after a cycle. On the
contrary we show that the associated dynamical phase is completely
insensitive to the assumed lack of symmetry of the trap. We thus
succeed in proposing simple and realizable appropriate initial
conditions which after a cycle do possess quantum coherences
transparently related to the non dynamical phase accumulation and
then to the anisotropy of the trap.

The main result of this paper is the idea of exploiting such a
deep geometrical-dynamical connection to propose, at least in
principle, a scheme to estimate quantitatively and qualitatively
the absence of symmetry in the trap.

To this end we present a motivated choice of a simple and
physically transparent observable whose mean value manifests
macroscopic differences if referred to an isotropic or not
oscillator.

Our analysis does not incorporate from the very beginning possible
sources of decoherence, and thus we wish to conclude discussing
the possibility of observing the reported anisotropy effects in
laboratory.

Condition (\ref{cap4_cond_ad}) for the adiabaticity of the time
evolution of the system imposes limitations to the square modulus
of the ratio between the anisotropy degree $\Delta\nu$ and the
coupling constant $\lambda$. Assuming accordingly
$\left|\Delta\nu/\lambda\right|^2\approx0,1$ immediately yields
$\Delta\nu_{\mbox{max}}\approx\lambda/3$.

In experimental works on ions in Paul traps \cite{meekhof},
testing an interaction described by a Jaynes-Cummings hamiltonian,
it has been seen that the effects of decoherence effectively
appear after about ten Rabi oscillations, i.e. the coherence time
$\tau$ of the ion  vibrational motion is about one tenth of the
coupling constant of the interaction hamiltonian. If we suppose
that this relation is also valid in the experimental scheme
proposed by us, i.e. $1/\tau\sim\lambda/10$, the time
$T=4\pi/\Delta\nu$ taken to complete an adiabatic cyclic evolution
is of the order of $\tau$. This means that, in such a scenario,
possible decoherence effects would be still weak enough after a
cycle to be negligible within a first approximation.

\end{document}